# A user-operator assignment game with heterogeneous user groups for empirical evaluation of a microtransit service in Luxembourg

Tai-Yu Ma*, Joseph Y.J. Chow, Sylvain Klein, Ziyi Ma

*Corresponding author

## Abstract

We tackle the problem of evaluating the impact of different operation policies on the performance of a microtransit service. This study is the first empirical application using the stable matching modeling framework to evaluate different operation cost allocation and pricing mechanisms on microtransit service. We extend the deterministic stable matching model to a stochastic reliability-based one to consider user's heterogeneous perceptions of utility on the service routes. The proposed model is applied to the evaluation of Kussbus microtransit service in Luxembourg. We found that the current Kussbus operation is not a stable outcome. By reducing their route operating costs of 50%, it is expected to increase the ridership of 10%. If Kussbus can reduce in-vehicle travel time on their own by 20%, they can significantly increase profit several folds from the baseline.

**Keywords**: Microtransit; On-demand mobility; Stable matching; Assignment game

## 1. Introduction

On-demand mobility (MOD) service has been promoted as an effective alternative to reduce traffic congestion and $CO_2$ emissions in many countries (Murphy and Feigon, 2016). A range of such MOD services include microtransit, ridesharing, paratransit, taxi, and ride-hailing, etc. have been successfully deployed in many cities with different service requirements and operation policies (Kwoka-Coleman, 2017; Metro magazine, 2019). As rural areas have low accessibility to public transport service, a microtransit system presents a good potential to compensate for this gap and reduce personal car use. Microtransit is any shared public or private sector transportation service that offers fixed or dynamically allocated routes and schedules *in response* to individual or aggregate consumer demand, using smaller vehicles and capitalizing on widespread mobile GPS and internet connectivity (see Volinski, 2019; Chow et al., 2020).

Although microtransit service can overcome the shortage of fixed-route public transit service, one of the main issues remains its high operating cost, generally much exceeding its revenue from ticketing and require government subsidy. A mix of successful ventures like Via and MaaS Global along with failed microtransit services like Kutsuplus (Haglund et al., 2019), Bridj (Bliss, 2017), and Chariot (Hawkins, 2019) show the importance of operating cost allocation decisions for a sustainable service operation. Service planning should consider travellers' choice preferences as well as operators' cost allocation policy to predict and evaluate ridership on the service network. For example, cost allocation policies may include fare prices, stop locations (which trade-off with access time), or frequency setting (which trade-off with wait time), which all involve distributing generalized travel costs that are transferable between users and operators.

Evaluation of platforms that support multiple operators, including microtransit services, requires models that capture both travellers' and operators' choices. There are a few methods that can address this requirement, but not without caveats. Bilevel network design models (Zhou, Lam, and Heydecker 2005) have been used to model markets of multiple transit operators as a generalized Nash equilibrium between operators. The interaction of travellers and operators can also be achieved by dynamic systems simulation via day-to-day adjustment (Djavadian and Chow, 2017a, b) toward the same noncooperative equilibrium. These models find one equilibrium between multiple operators; however, equilibria can be non-unique and dependent on a governing platform's (or public agency acting as one) mechanism design. In other words, a noncooperative game framework between operators limits the design considerations for a market that can incorporate subsidies, fare bundling, transfer locations, etc., that would involve more flexible transfer of utility between operators and travellers.

Assignment games (Shapley and Shubik, 1971; Sotomayor, 1992) are a form of transferable utility (TU) stable matching model, also called a TU-game, that outputs the set of stable outcomes corresponding to an optimal assignment. Whereas only a profit-maximizing objective is allowed in a noncooperative game in the bilevel network design problems, assignment games allow for a range of outcomes that can include both welfare-maximizing objectives as well, or for any mechanism that lies in between. In assignment games, these two opposing outcomes are called the buyer-optimal outcome (welfare maximizing) and seller-optimal outcome (profit maximizing). This allows platforms in which different operators may seeks different objectives (including having both public and private operators or operators seeking a hybrid objective). Travellers receive a net utility from using the service and transfer the cost (ticket price) as a benefit to the operator. The platform's assignment game is formulated as two subproblems: a matching or assignment problem in which travellers and operators are optimally assigned to each other, and a stable outcome problem that ensures that the matching has sufficient incentives (non-zero profit on each side) to participate.

Rasulkhani and Chow (2019) proposed an assignment game for modelling a platform that includes a set of capacitated operator-routes and a set of travellers. Travellers' preference, generalized travel cost, and operators' routing cost and service design options can be explicitly considered. This approach allows the platform to evaluate the impact of different operating policies on ridership. While subsequent studies have extended the work to include generalized multimodal trips (Pantelidis, Chow, and Rasulkhani, 2019), no empirical study has been conducted with this methodology yet.

The contribution of this study is threefold. We study a microtransit service as a platform hosting a set of vehicle-operators serving travellers in which data is available. To make use of such data, we first propose a stochastic variant of the user-operator stable outcome subproblem to match users and a set of service lines with capacity constraints. This model takes into account the heterogeneous nature of users' perceived travel utility, resulting in a probabilistic stable operation cost allocation outcome to design ticket price and ridership forecasting. We show the stochastic stable outcome problem corresponds to that heterogeneous matching subproblem. The stochastic variant is necessary to incorporate real data that exhibits heterogeneous behaviour. Second, we develop the methodology to estimate the model parameters and calibrate them to evaluate an operator's service policy. Third and primarily, we apply the proposed approach to an empirical study of a microtransit service, namely Kussbus[1], in Luxembourg and its French- and Belgium-side border area using real data shared by the company UFT (Utopian Future Technologies S.A.). We conduct a sensitivity analysis to investigate the impact of route cost, in-vehicle travel time and access distance to bus stops on ticket prices, ridership and operator's profit. The results support the new approach and tool to evaluate different operating and cost allocation policies for operators.

## 2. Methodology

### 2.1. Stable matching model with heterogeneous user groups

The stable matching problem has been studied since 1960 (Gale and Shapley, 1962) to determine an optimal matching involving multiple participants on a two-sided matching market. Early studies (Shapley and Shubik, 1971; Sotomayor, 1992) formulate the problem as an assignment game to find an optimal matching to form coalitions between buyers and sellers along with feasible transfers of utility between participants such that no participant has incentive to break the coalition. The assignment game approach has been applied in collaborative transportation problems to set up stable cost/profit allocation mechanism to form profitable collaboration between participants (Agarwal and Ergun, 2010; Verdonck et al., 2016; Schulte et al., 2019). In the basic form of the assignment game (Sotomayor, 1992), two distinct set of players, i.e. sellers J and buyers I, are considered. In such a setting, a seller j provides service with a cost $c_j$ while a buyer $i$ pays a price $p$ and receives a utility $U_{ij}$. The payoffs generated from a seller-buyer matching is $a_{ij} = max(0, U_{ij} - c_j)$ where buyer $i$ gains a net utility $u_i = U_{ij} - p$, and seller $j$ a net profit of $v_j = p - c_j$. An assignment game is a type of transferable utility game (TU game), which belongs in the set of cooperative games involving stable matching. The basic form can be extended to set up a cost allocation mechanism to evaluate various operation policies of capacitated route mobility services (Rasulkhani and Chow, 2019) and collaborative Mobility-as-a-Service platforms (Pantelidis, Chow and Rasulkhani, 2019).

---

[1] https://kussbus.lu/

In the context of user-operator assignment game for a microtransit service platform, the problem considers a set of users, $s \in S$, to be assigned to a set of routes, $r \in R$, provided by one or multiple operators. The problem is a many-to-one assignment game in which one route can be matched to multiple users and one user can match to only one route (see an illustrative example in Figure 1). **The decision-maker is the platform (not necessarily a public agency) in which each microtransit vehicle run is an operator**. The stable matching approach first considers optimal assignment of users' ride requests and operator's service routes and then determines a route cost allocation and pricing mechanism to ensure the assignment is stable. Each route is a sequence of stops visited by one or more vehicles on which line (for multiple vehicles) or vehicle (for single vehicle) capacity constraints need to be satisfied. We consider each user a buyer, and each route a seller with a selling price for using that portion of the service route. When users are assigned to routes, users pay a respective ticket price and gain a payoff upon trip completion, while the seller gains a profit as the revenue received from a user reduced by the cost allocated to the supply of that portion of routes. The objective of the assignment game is to find a seller-buyer matching/assignment such that a total generalized payoff is maximized. The outcome of the model is route flows as well as stable cost/profit allocation outcome, i.e. user payoff and operator profit profiles. This is the same logic as Shapley and Shubik (1971) as a transferable utility game involving coalition formation between the operators (as sellers) and users (as buyers). The cost allocation outcome can be used to design ticket prices and other travel disutilities (e.g. wait or access times due to matching algorithms which impact the total payoff available for cost allocation) and evaluate their impact on ridership and operator revenue. The reader is referred to Rasulkhani and Chow (2019) for a more detailed description of the model properties.

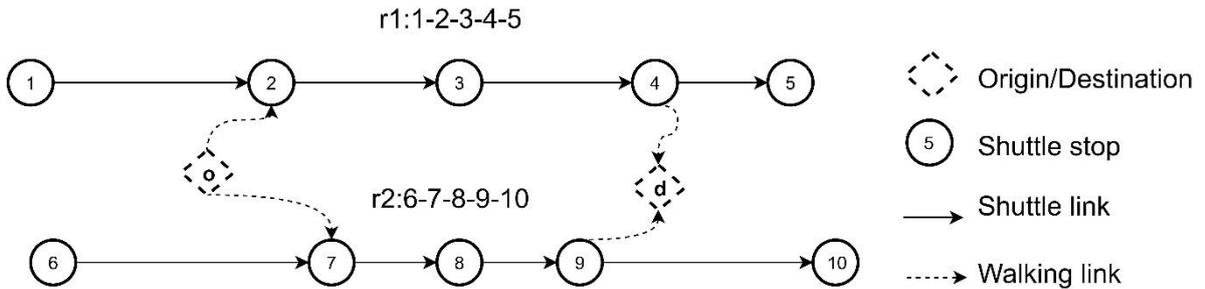

**Figure 1**. Example of one user and two routes (r1 and r2). A user's generalized travel cost includes a door-to-door travel cost as the weighted sum of walking time, waiting time, in-vehicle travel time, and ticket price paid to the operator.

The stable matching model is as two subproblems. First, an optimal user-route matching problem (P1, below) determines user-route matches that maximize total generated payoff. The output of P1 is a set of matched user-route flows on the operator's service network. Second, given the matching result of P1, a stable outcome problem (P2, described later) is used to determine the stable outcome space corresponding to the assignment in which operators and users have no incentive to switch (for users this might involve switching to other service routes or a dummy route for no travel or an option external to the market; for operators this involves matching to other users). The output of P2 is the profile of net payoffs for users and routes of operators, i.e. $(u_s, v_r)$. **The overall problem is a transferable utility game** where P1 finds the best matches and P2 provides a set of stable cost allocations from which the platform can choose. The utility in P1 is the total gain from which the benefits of both the operator and the users split the profits if they match successfully. Note we **measure all transferable utility, payoff and profit in monetary units (euros)** in this study.

**P1: User-route matching model**

$$\max \sum_{s \in S} \sum_{r \in R} a_{sr} x_{sr} \quad (1)$$

s.t.

$$\sum_{r \in R} x_{sr} \leq d_s, \forall s \in S \setminus \{k\} \quad (2)$$

$$\sum_{s \in S \setminus \{k\}} \delta_{asr} x_{sr} \leq q_r, \forall a \in A_r, r \in R \quad (3)$$

$$\sum_{s \in S \setminus \{k\}} x_{sr} \leq M(1 - x_{kr}), \forall r \in R \quad (4)$$

$$x_{sr} \in \{0, \mathbb{Z}_+\}, \forall s \in S \setminus \{k\}, \forall r \in R \quad (5)$$

$$x_{kr} \in \{0,1\} \quad (6)$$

The objective function (1) maximizes total payoff gains form the matching. $a_{sr}$ is the net payoff of a user($s$)-route($r$) match. The payoff gained by a user $s$ for matching with route r is $a_{sr} = \max(0, U_{sr} - t_{sr})$, where $U_{sr}$ is the utility gain for user s using route r, $t_{sr}$ is the generalized travel cost for user-route pair $(s, r)$. The latter parameter $t_{sr}$ can be tuned to account for many different policy or algorithm designs as well as scenario settings. For example, one can specify $t_{sr} = t_{sr,IV} + b_1 t_{sr,wait} + b_2 t_{sr,access}$ as three terms for in-vehicle time (IV), waiting, and access with corresponding coefficients $b_1, b_2$. In that case, an operator interested in evaluating a new matching algorithm that would on average increase access time for users but reduce wait time and in-vehicle time as well as operating cost $C_r$ of route $r$ can use this model to compare the effect of the operating designs. A city agency wanting to measure the effect of increased travel times due to added congestion on the roads can increase the in-vehicle time to see how that impacts the assignment game outcomes.

Equation (2) states for any user $s$ the summation of flows over routes cannot exceed its demand $d_s$. Equation (3) states assigned user flow on any route needs to satisfy corresponding route capacity constraint $q_r$ (passengers per hour). $\delta_{asr}$ is an indicator being 1 if arc $a$ is used by user $s$ for route $r$ and 0 otherwise. The dummy user $k$ of not matching with any route is set as a reference alternative, generally with a utility of 0. This assumes that the market has no other travel options outside the system that provides travel utility for matching (i.e. a closed market as opposed to an open market or submarket controlled with a platform). Equation (4) ensures that a route is only matched when its total payoff exceeds a threshold cost; for private operators with no subsidy this would be setting $a_{kr} = C_r$. $M$ is a big positive number. Equation (5) ensures that the decision variable $x_{sr}$ is a non-negative integer which is a more generalized case where $x_{sr}$ can be larger than 1. Only $x_{kr}$ (dummy user for inactive routes) (Equation (6)) needs to be binary.

Departing from Rasulkhani and Chow (2019), we make the following modification to the model to allow us to forecast utility from route-level or user-level attributes. In the original model, each user group $s \in S$, typically representing an origin-destination (OD) pair, is assumed homogeneous. In this study, we assume users for a particular OD group are heterogeneous: the utility $U_{sr}$ is an independent random variable composed of a deterministic part $V_{sr}$ and an unobserved part $\varepsilon_{sr}$ as Equation (7).

$$U_{sr} = V_{sr} + \varepsilon_{sr} \qquad (7)$$

where $V_{sr}$ is the mean utility gain of a trip and $\varepsilon_{sr}$ is a random utility term that follows a Normal distribution with mean 0 and standard deviation $\sigma$. Note that one can extend this assumption by considering more various distributions. Given $U_{sr}$ is probabilistic, so is $a_{sr}$. Objective (1) is modified to Eq. (8) to reflect the optimization of the expected value of $a_{sr}$.

$$max\ \mathbb{E}\left[\sum_{s\in S}\sum_{r\in R} a_{sr} x_{sr}\right] = max\left[\sum_{s\in S}\sum_{r\in R} \max(0, V_{sr} - t_{sr}) x_{sr}\right] \qquad (8)$$

Let us call the heterogeneous form of P1 where the objective function (1) is replaced with Eq. (8) as P1H.

**P2: User-operator stable outcome model**

Per Rasulkhani and Chow (2019), the stable outcome model is specified as follows in Equations (9) – (14).

$$max\ Z \qquad (9)$$

s.t.

$$\sum_{s\in G(r,x)} u_s + v_r \geq \sum_{s\in G(r,x)} a_{sr} - C_r, \forall G(r,x)\ and\ r \in R \qquad (10)$$

$$\sum_{s\in S(r,x)} u_s + v_r = \sum_{s\in S(r,x)} a_{sr} - C_r, \forall r \in R^* \qquad (11)$$

$$v_r = 0, \forall r \in R\setminus R^* \qquad (12)$$

$$u_s = 0, \quad if\ s \in \bar{S} = \{s|\sum_{r\in R} x_{sr} = 0\} \qquad (13)$$

$$u_s \geq 0, v_r \geq 0, \forall r \in R^* \qquad (14)$$

Equation (9) is the objective function to be maximized. Based on the design objective, one can set $Z = \sum_{s\in S} u_s$ to maximize total utility gain of users ($u_s$), which would be a vertex of interest to public agencies. Its solution, if any, is a buyer-optimal cost allocation outcome. Alternative, if we aim to maximize total profit gain of operators ($v_r$), the objective function becomes $Z = \sum_{r\in R} v_r$. The optimal solution is a seller-optimal outcome. If no cost allocation mechanism is being evaluated, one can solve the stable outcome problem twice, once for buyer-optimal and again for seller-optimal objective, to obtain the vertices for the full range of stable outcomes from which the platform can select one. Since the problem is convex (a set of linear constraints), the prices based on convex combinations of the two vertices would all be stable as well (Rasulkhani and Chow, 2019). Equation (10) ensures the stable condition for which no user would have a better payoff other than the current assignment. $G(r,x)$ is the group of users which can be feasibly assigned on route $r$ given the solutions $x$ of P1. In the case of an operator owning multiple routes, constraint (10) is only applied to routes not owned by that operator. In other words, in the case of a centralized operator where costs can freely transfer between routes, constraint (10) would be relaxed. For the Kussbus case study we assume routes do not freely transfer costs between each other.

The feasibility constraints are verified when $G(r,x)$ satisfies Equation (3). For example, consider a matching outcome $x$ assigns users $\{s_1, s_2, s_3\}$ to a route $r$. The set of group users $G(r,x)$ is the union of subsets of users from $\{s_1, s_2, s_3\}$, i.e. $\{\{s_1\}, \{s_2\}, \{s_3\}, \{s_1, s_2\}, \{s_1, s_3\}, \{s_2, s_3\}, \{s_1, s_2, s_3\}\}$. Equation (11-13) are the feasibility conditions where $R^*$ is the subset of routes with at least one matched user. Equation (11) split the total utility gain between the user and the operator. $S(r, x)$ is the set of users matching route $r$, given an optimal matching solution $x$ obtained by P1. Equation (14) ensures the decision variable $u_s$ and $v_r$ are non-negative continuous variables. We call $\{(u, v); x\}$ a cost allocation outcome given an optimal matching $x$. The cost allocation outcome is the list of payoffs and profits for users and routes.

New in this study, Equation (10) and (11) are modified to stochastic constraints because of the presence of a stochastic $a_{sr}$. By introducing Equation (7) in (10), Equation (10) becomes Equation (15).

$$\sum_{s \in G(r,x)} u_s + v_r + C_r - \sum_{s \in G(r,x)} max(0, V_{sr} - t_{sr}) \geq \sum_{s \in G(r,x)} \varepsilon_{sr}, \quad \forall G(r,x), r \in R \quad (15)$$

The constraint represents a heterogeneous population, which is not the same as the "stochastic stability" condition (see Fernández et al., 2002; Sawa, 2014; Klaus and Newton, 2016) in evolutionary games with random perturbations. Instead, under heterogeneity the stability condition can only be guaranteed for a portion of the population. Consider the concept of $\alpha$-stability in Definition 1.

**Definition 1**. An outcome $\{(u, v); x\}$ for a population of heterogeneous user groups $S$ is **$\alpha$-stable** if $(1 - \alpha)$ of each user group $s \in S$ meets the stability condition in Eq. (10). When $\alpha = 0.50$ the stability condition simplifies back to the deterministic case of Rasulkhani and Chow (2019).

For example, a 0.05-stable outcome for a heterogeneous user assignment game implies cost allocations that can only guarantee stability for 95% of the users. Then Equation (15) can be expressed deterministically as a chance constraint (16).

$$\Phi\left[\frac{\left(\sum_{s \in G(r,x)} u_s + v_r + C_r - \sum_{s \in G(r,x)} max(0, V_{sr} - t_{sr})\right)}{|G(r,x)|}\right] \geq 1 - \alpha, \quad \forall G(r,x), r \in R \quad (16)$$

where $\Phi(z) = \Pr(Z \leq z)$ is the cumulative density function of Z.

Equation (16) is a nonlinear constraint which can be transformed to a linear inequality in Equation (17) per Shapiro, Dentcheva, and Ruszczyński (2009).

$$\sum_{s \in G(r,x)} u_s + v_r + C_r - \sum_{s \in G(r,x)} max(0, V_{sr} - t_{sr}) \geq Z_{1-\alpha}\sigma', \quad \forall G(r,x), r \in R \quad (17)$$

with new deviation $\sigma' = \sqrt{|G(r,x)|\sigma^2}$. Eq. (11) can be correspondingly converted to Eq. (18).

$$\sum_{s \in S(r,x)} u_s + v_r + C_r - \sum_{s \in S(r,x)} max(0, V_{sr} - t_{sr}) = \mathbb{E}\left(\sum_{s \in S(r,x)} \varepsilon_{sr}\right) = 0, \quad \forall r \in R^* \quad (18)$$

Let us call P2 with Eq. (10) – (11) replaced with Eq. (17) and (18) as P2H. P2H is a linear programming problem and can be solved efficiently by the simplex algorithm or interior-point algorithm using existing commercial solvers.

Given the $\alpha$-stability outcome $\{(u, v); x\}_\alpha$, we can determinate ticket prices for user $s$ using route $r$ as shown in Equation (19).

$$p_{sr} = v_{sr} + c_{sr}, \forall r \in R, \forall s \in S\backslash\{k\} \quad (19)$$

where $v_{sr}$ is the profit gained of operating route r from user s. $c_{sr}$ is the cost of operating route r to be transferred to user s. We have $\sum_{s\in S(r,x)} c_{sr} = C_r$ and $\sum_{s\in S(r,x)} v_{sr} = v_r, \forall r \in R$.

Operators may determine the ticket prices based on a preferred policy (e.g. equal-share or cost-based share policy) given the stable cost allocation outcome. Given a user-route matching outcome in P1H and a value $\alpha$ for the platform, one can solve the P2H problem to obtain buyer- and seller-optimal outcomes. An example is shown in Section 2.2.

## 2.2. Illustrative example

We illustrate the stable matching model with heterogeneous user groups by a simple 4-node transit route example drawn from Rasulkhani and Chow (2019) (see Figure 2). Travel time from node $i$ to node $j$ is shown on Figure 2. The operation cost of a route is assumed as a function of number of its links, defined as $C_r = 4.5 + 0.5|A_r|$, where $A_r$ is the set of arcs on route $r$. We assume all possible routes can be enumerated resulting in 52 possible routes. A total of 60 users (demand) is generated with 10 users for each user group, i.e. $d_s = 10, \forall s \in S = \{s_1, s_2, s_3, s_4, s_5, s_6\} = \{(1,2),(1,3),(2,3),(3,2),(4,1),(4,2)\}$. The line capacity $q_r, \forall r \in R$ is set to 6. The utility of trip is assumed as $U_{rs} = V_s + \varepsilon_s, \forall s \in S$, with $V_s = 20$ for $s_1 - s_3$, and $V_s = 25$ for $s_4 - s_6$. $\varepsilon_s \sim \mathbb{N}(0, \sigma_s)$ with $\sigma_s = 1$, 2, and 3 for $\{s_1, s_4\}, \{s_2, s_5\}$, and $\{s_3, s_6\}$, respectively.

We first solve P1H to obtain optimal user-route flows $x_{sr}$. Then we solve P2H using the results of P1H to determine ticket prices $p_{sr}$. Each route $r$ should charge each user group $s$ under different reliability $(1-\alpha)$ to meet the stable condition (Eq. (17)). Two design objectives (buyer-optimal ($Z = \sum_{s\in S} u_s$) and seller-optimal ($Z = \sum_{r\in R} v_r$)) are considered. We use MATLAB *intlinprog* and *linprog* solvers to solve P1H and P2H problems, respectively.

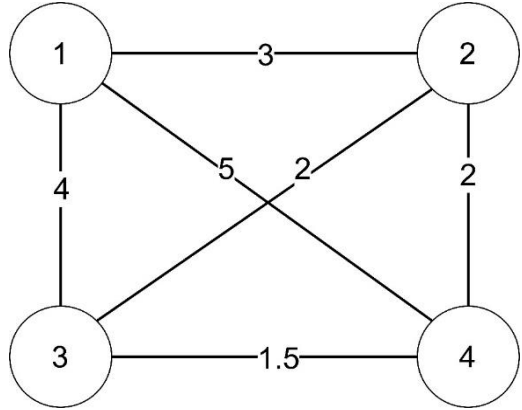

**Figure 2.** A simple 4-node network example.

The optimal user-route matching result is shown in Table 1. The optimal objective function value is 1137. All users are matched with 6 routes where route (4-2) has 4 users and route (4-1-3-2) has 17 users. Given the user-route matches, we set up the ticket price based on buyer-/seller- optimal objectives with different level of stabilities $\alpha \in \{0.5, 0.4, 0.3, 0.2, 0.1\}$. The results are shown on Tables 2 and Table 3. Under the buyer-optimal objective, the total payoffs of users are almost the same with $Z^* = 113.7$ (the last line of Table 2) given different values of $\alpha$. The lack of change is because the operator fares are all pushed to the minimum allowed. However, under the seller-optimal objective, increasing (1- $\alpha$) results in lower ticket price $p_{sr}$ and lower total payoff of operator (Table 2). When there's heterogeneity, smaller values of $\alpha$ require a higher percent of the population to be satisfied, resulting in less room to maximize fare price and profit. As a result, the convex stable outcome region between buyer- and seller-optimal spaces shrinks (and even collapses to a unique value for Route 6 for $\alpha \leq 0.3$).

**Table 1. Result of user-route matches on the illustrative example.**

| r Route number | Links of route | Cost of route | User group $(o,d)$ | | | | | | Total |
|---|---|---|---|---|---|---|---|---|---|
| | | | (1,2) | (1,3) | (2,3) | (3,2) | (4,1) | (4,2) | |
| 6 | 4-2 | 5 | 0 | 0 | 0 | 0 | 0 | 4 | 4 |
| 7 | 1-3-2 | 5.5 | 0 | 5 | 0 | 4 | 0 | 0 | 9 |
| 9 | 1-2-3 | 5.5 | 6 | 0 | 4 | 0 | 0 | 0 | 10 |
| 25 | 4-1-2 | 5.5 | 4 | 0 | 0 | 0 | 4 | 0 | 8 |
| 28 | 4-2-3 | 5.5 | 0 | 0 | 6 | 0 | 0 | 6 | 12 |
| 49 | 4-1-3-2 | 6 | 0 | 5 | 0 | 6 | 6 | 0 | 17 |
| Total | | | 10 | 10 | 10 | 10 | 10 | 10 | 60 |

**Table 2. Ticket prices in buyer-optimal and seller-optimal allocation mechanisms under different levels of stability.**

| Route number | Buyer-optimal | | | | | Seller-optimal | | | | |
|---|---|---|---|---|---|---|---|---|---|---|
| | $\alpha$ | | | | | $\alpha$ | | | | |
| | 0.5 | 0.4 | 0.3 | 0.2 | 0.1 | 0.5 | 0.4 | 0.3 | 0.2 | 0.1 |
| 6 | 1.25 | 1.25 | 1.25 | 1.25 | - | 4.00 | 3.71 | 1.25 | 1.25 | - |
| 7 | 0.61 | 0.61 | 0.61 | 0.61 | - | 1.67 | 1.36 | 1.99 | 1.35 | - |
| 9 | 0.55 | 0.55 | 0.55 | 0.69 | - | 3.30 | 2.77 | 1.97 | 0.81 | - |
| 25 | 0.69 | 0.69 | 0.69 | 0.69 | - | 2.25 | 2.25 | 1.64 | 0.81 | - |
| 28 | 0.46 | 0.46 | 0.46 | 0.46 | - | 2.75 | 1.95 | 0.63 | 0.63 | - |
| 49 | 0.35 | 0.35 | 0.35 | 0.42 | - | 1.18 | 0.93 | 1.43 | 1.15 | - |
| $Z^*$ | 113.7 | 113.7 | 113.7 | 113.4 | - | 102.0 | 79.0 | 54.4 | 25.7 | - |

Remark: - : no solution

**Table 3. Operator's profit on different routes in buyer-optimal and seller-optimal allocation mechanisms under different level of stability.**

| Route number | Buyer-optimal | | | | | Seller-optimal | | | | |
|---|---|---|---|---|---|---|---|---|---|---|
| | $\alpha$ | | | | | $\alpha$ | | | | |
| | 0.5 | 0.4 | 0.3 | 0.2 | 0.1 | 0.5 | 0.4 | 0.3 | 0.2 | 0.1 |
| 6 | 0.0 | 0.0 | 0.0 | 0.0 | - | 11.0 | 9.8 | 0.0 | 0.0 | - |
| 7 | 0.0 | 0.0 | 0.0 | 0.0 | - | 9.5 | 6.7 | 12.4 | 6.6 | - |
| 9 | 0.0 | 0.0 | 0.0 | 1.4 | - | 27.5 | 22.2 | 14.2 | 2.6 | - |
| 25 | 0.0 | 0.0 | 0.0 | 0.0 | - | 12.5 | 12.5 | 7.6 | 1.0 | - |
| 28 | 0.0 | 0.0 | 0.0 | 0.0 | - | 27.5 | 17.9 | 2.0 | 2.0 | - |
| 49 | 0.0 | 0.0 | 0.0 | 1.1 | - | 14.0 | 9.9 | 18.3 | 13.5 | - |

Remark: - : no solution

## 3. Stable matching application case study

We present a methodology to estimate and calibrate the model parameters for the stochastic assignment game model and evaluate different service design such as access time, ride time, and paid fare on the operator's revenue and the ridership. More precisely, it aims to respond to the following research questions:

- Based on the estimated utility parameters and the characteristics of the routes, the model predicts a stable outcome range for user ticket prices for a given reliability measure $\alpha$. Having the individual ride observations, how should one calibrate $\alpha$ if the objective is to maximize matches between predicted vehicle-route flow and the observed data?
- What is the impact of different pricing policies on the ridership and operator's profit?
- If Kussbus should focus on one area to improve upon (i.e. reduction in operating cost, reduction in access time, or in-vehicle time), which should they focus on to increase ridership and what would be the resulting impact on its net profit?

### 3.1. Data and case study setting

Kussbus Smart shuttle service (https://kussbus.lu/en/how-it-works.html) is a first microtransit service operating in Luxembourg and its border area. The service was provided by the Utopian Future Technologies S.A.(UFT) from April 2018 to March 2019. Like most microtransit systems, users use dedicated Smartphone applications to book a ride in advance with desired origin, destination and pickup time as input. Service routes are flexible to meet maximum access distance constraint. Routes are generated in a way that we assume users need to walk from/to the origin/destination to/from shuttle stops given a pre-defined threshold (i.e. around one kilometer). The service started operating between the Arlon region in Belgium and the Kirchberg district of Luxembourg City on 04/25/2018 and a second line started on 09/24/2018 between Thionville region (France) and Kirchberg district. Both service areas are highly congested on road networks due to high car use during morning and afternoon peak hours (Rifkin et al., 2016).

The empirical ride data was provided by the operator for the period from 4/25/2018 to 10/10/2018. A total of 3258 trips (rides) were collected. Each ride contains the following information: booking date and time, pickup time and drop-off time, pickup and drop-off locations, pickup and drop-off stops, walking distance between stops and origins/destinations, origin-destination pairs of users, and fare. Any abnormal trips (e.g. trip duration less or equal to 5 minutes) were removed. As a result, a total of 3010 trips were used for this study.

We summarize the characteristics of Kussbus service as follows. More detail about the operation policy and characteristics of Kussbus service can be found at: https://uft.lu/en/news/references/kussbus.

- Service areas: two service areas: a.) Arlon region (Belgium) $<->$ Kirchberg district (Luxembourg City), and b.) Thionville region (France) $<->$ Kirchberg district.
- Operating hours: From 05:30 to 09:30 and from 16:00 to 19:00 from Monday to Friday.
- Vehicle capacity: vehicle capacity differs from 7-seater, 16-seater, and 19-seater.
- Booking and ticket price: users need to book a ride by the dedicated Smartphone application. First 6 trips are free, and then the unit ticket price is around 5 euros per trip.
- Vehicle routing policy: vehicle routes are scheduled based on pre-booked customer requests on previous days. Late-requests could be accepted under certain operational constraints.

The entire study period of Kussbus riding data contains 235 commuting periods in the morning or afternoon from April to October 2018. The operator's routes are generated beforehand based on the observed routes in the data. We solve P1 and P2 under a multi-period, static setting.

There are 429 possible routes observed from Arlon to Kirchberg (see Figure 3) and 449 in the reverse direction. From Thionville to Kirchberg there are 52 routes (see Figure 4) with 50 routes in the reverse direction. The average operation costs takes into account driver and fuel costs. For the operating cost of route (i.e. vehicle-route), it is estimated as the average operating cost per kilometer travelled multiplied by travel distance. Route travel time is estimated by Google Maps API during corresponding peak hour traffic conditions. Table 4 reports the characteristics of Kussbus service and relevant parameter settings for the case study.

Table 4. Kussbus service characteristics and parameters settings.

| Attribute | Value | Attribute | Value |
|---|---|---|---|
| Number of trips | 3010 | User's maximum waiting time at stop | 10 minutes |
| Value of in-vehicle time (VOT) (euro/hour)* | 24.21 | Capacity of vehicles | 7, 16 and 19 passenger seats |
| Walking speed | 5km/hr | Average route cost | 61.0 euros |
| Average route distance | 46.5 km | Average travel distance of users | 43.0 km |

* based on the estimation in this study.

We calibrate users' utility (Equation (7)) to fit observed user-route matches. For this purpose, we divide the data into a training dataset (first 80% rides (213 commuting periods)) and a test dataset (remaining 20% rides (remaining 22 periods)). The calibration consists of two steps. The first step consists of estimating the value of in-vehicle travel time (VOT). The estimated VOT can then be used to estimate users' generalized travel costs. The second step consists of calibrating the users' utility values to fit observed user-route matches (i.e. user-used route pair) over the studied period. We use the commercial solver *intlinprog* of MATLAB to solve the P1 and P2 problems based on a Dell Latitude E5470 laptop with win64 OS, Intel i5-6300U CPU, 2 Cores and 8GB memory.

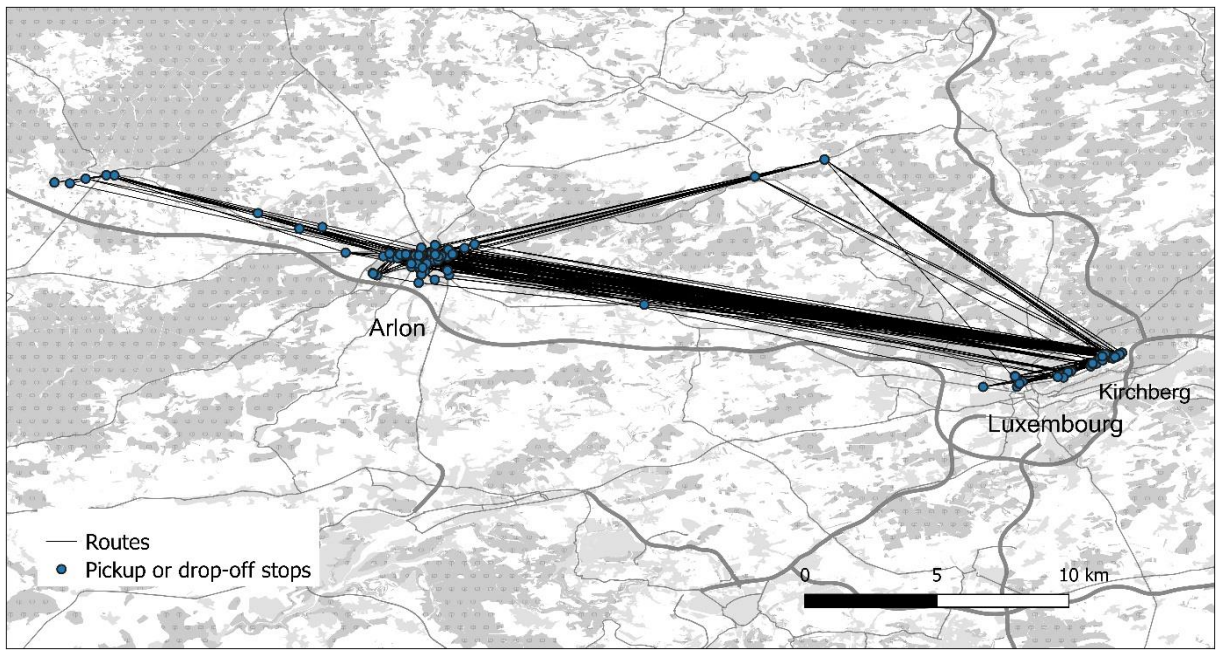

**Figure 3**. Kussbus operating routes from Arlon to Luxembourg City.

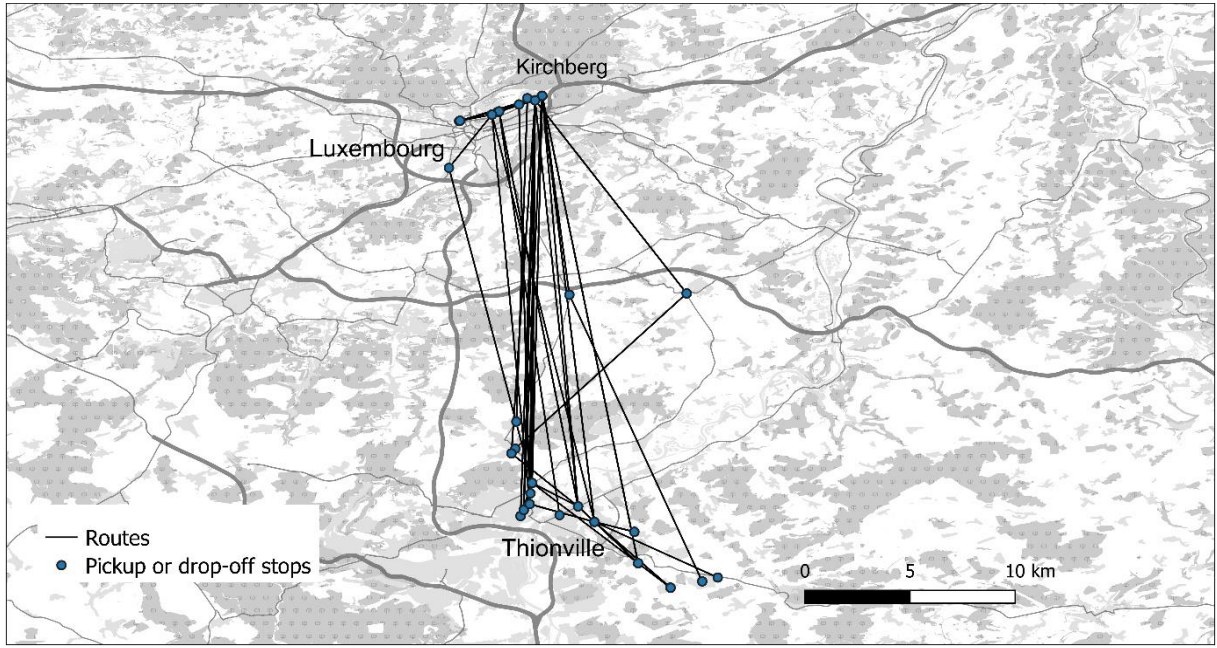

**Figure 4**. Kussbus operating routes from Thionville to Luxembourg City.

*3.2. VOT estimation*

To estimate the VOT for commuting trips in the study area, we use a mobility survey conducted in October-November 2012 for the EU officials and temporary employees working in the European institutions at the Kirchberg district of Luxembourg. The survey contains samples living in Luxembourg and its French, Belgium and Germany border areas, which perfectly matches Kussbus's service area. A total of 370 valid samples (individuals) were collected in which there are 131 individuals from the European Investment Bank (~6.2% of total staff in 2012) and 239 individuals from the Court of Justice of the European Union (~11.2% of total staff). After a data cleaning process, a total of 309 individuals' commuting trip data were used for the analysis. The spatial distribution of respondents' residential locations appeared as Luxembourg (78.3%), France (9.4%), Belgium (7.8%) and Germany (4.5%). Note that Belgium employees live mainly in Arlon (45.8% of Belgium employees). French employees live mainly in Thionville, Hettange-Grande, and Yutz (44.8% of French employees). As only 5% of the sample use 'walk' and 'bicycle' as commuting mode, these samples are excluded from the analysis. We focuses on bimodal (car and public transport) mode choice case, which is consistent with the current mode share in the study area ("Luxmobil" survey, 2017).

Based on previous studies (Gerber et al., 2017; Ma, 2015, Ma, Chow, and Xu 2017), explanatory variables for mode choice include individual-specific socio-demographic variables (gender, age, presence of children etc.), and alternative-specific variables (i.e. travel time and travel cost etc.). Two discrete choice models are specified: a multinomial probit model (MP) and a mixed logit model (ML). The mixed logit model allows random preference coefficient specification to capture travelers' preference heterogeneity (Train 2003). As no convergent estimation results were obtained for the mixed logit model, we only report the estimation results of the MP model in Table 5. The first model (MP-1) considers relevant socio-demographic variables and mode-specific variables. The second model (MP-2) further incorporates spatial-specific variables related to the municipality of respondents' residential locations. The likelihood ratio test shows the MP-2 outperforms the MP-1 at a statistical significance level of 0.05 (Prob. > chi-square=0.0148). We retain the MP-2 model as the final selected model.

Regarding the estimated coefficients in the final model, the results are consistent. Travel time and travel cost have negative effects on individuals' choices on car use. Free parking at the workplace encourages individuals to use car. Similarly, season ticket subscriptions might be related to frequently public transport users who prefer public transport. Number of children and number of cars in the household positively influence individuals' preferences to use car as a commuting mode. This result might be explained by the convenience of using cars for pickup/drop-off needs when children are present in the household. Luxembourg residents have significant preference for using car as a commuting mode due to lower accessibility to public transport in rural area, and other reasons related to habits, social and cultural norm. The estimated VOT for the MP-2 is 24.21 euro/hour which is consistent with existing VOT studies related to Luxembourg's situation[2] (Wardman, Chintakayala, and de Jong., 2016).

**Table 5. Estimation results of the multinomial probit models with different model specifications.**

|  | MP-1 | | MP-2 | |
| --- | --- | --- | --- | --- |
| Variable | Coef. | Std. | Coef. | Std. |
| Travel time | -0.013 | 0.009 | -0.023* | 0.012 |
| Cost | -0.155*** | 0.060 | -0.057 | 0.072 |
| Free_parking | 0.589* | 0.349 | 0.608* | 0.354 |
| Season_ticket | -1.050*** | 0.249 | -1.025*** | 0.254 |
| Gender | -0.183 | 0.236 | -0.176 | 0.238 |
| Couple | -0.669** | 0.329 | -0.720** | 0.333 |
| Age34 | -0.377 | 0.411 | -0.297 | 0.417 |
| Age35_44 | -0.169 | 0.385 | -0.138 | 0.389 |
| Age45_54 | -0.711* | 0.398 | -0.722* | 0.405 |
| N_children | 0.329*** | 0.124 | 0.350*** | 0.127 |
| N_car | 1.193*** | 0.224 | 1.225*** | 0.230 |
| Flex_time | 0.014 | 0.320 | -0.091 | 0.331 |
| Res_lux |  |  | 1.711** | 0.767 |
| Res_fr |  |  | 0.340 | 0.858 |
| Res_be |  |  | 1.042 | 0.780 |
| Constant | -0.981* | 0.593 | -2.773*** | 1.012 |
| Number of individuals | 309 | | 309 | |
| Log-likelihood value at convergence | -161.65 | | -156.51 | |
| Degree of freedom | 13 | | 16 | |
| Prob. > chi-square | <0.000001 | | <0.000001 | |
| Pseudo $R^2$ | 0.2446 | | 0.2686 | |
| AIC | 349.3 | | 345.02 | |
| BIC | 397.8 | | 404.8 | |
| Adjusted Pseudo $R^2$ | 0.1885 | | 0.1985 | |
| Likelihood ratio test ( Prob. > chi-square) | <0.00001 | | 0.0148 (MP-2 vs MP-1) | |

Remark: *0.05 < $p$-value ≤ 0.1. **0.01 < $p$-value ≤ 0.05. ***$p$-value ≤ 0.01

*3.3. Utility calibration*

We calibrate the utility $U_{sr}$ using the first 80% training dataset to maximize the user-route matches between observation and model predicted results. As no available survey data is available to direct estimate user commuting trip utility, we approximate it as an equivalent door-to-door car-use generalized travel cost ($U_s^{car}$) plus a constant utility ($U_s^0$) to be calibrated (i.e. $U_{sr} = U_s^0 + U_s^{car} = U_s^0 + \hat{c}_s + \varepsilon_{sr}$). Note that $U_s^{car}$ represents the perceived cost of the reference mode, and $U_s^0$ is the differentiation value between car and Kussbus service (Breidert, 2005). We estimate users' car-use generalized travel cost as $VOT \times t_s + \bar{c}_{car} \times d_s$, where $t_s$ is a user's trip travel time from origin to destination and $d_s$ is the trip travel distance. $\bar{c}_{car}$ is the average cost per kilometer travelled by car estimated as 0.2534 euros/km by considering fuel cost, vehicle purchase cost and annual assurance cost, which is consistent with an existing study (Victoria Transport Policy Institute, 2009). Given user's origin and destination, we use Google's API to estimate $\hat{c}_s$ by considering realsitic road congestion effect given user's departure time. User's generalized travel cost $t_{sr}$ is estimated by considering walking time $T^{walk}$ to nearest shuttle stop, waiting time $T^{wait}$, and riding time $T^{ride}$ of trip, estimated as Equation (20).

---

[2] Wardman, Chintakayala, and de Jong (2016) estimated the values of time (€ per hour based on 2010 incomes and prices) for car commute is 18.06 (urban free flow) and 25.68 (urban congestion) in Luxembourg. For car business travel, it is 37.94 euros/hour in urban free flow situation and 53.95 euros/hour in urban congestion situation.

$$t_{sr} = \tau_1 T^{walk}_{Ov_1} + \tau_2 T^{wait}_{v_1} + \tau_3 T^{ride}_{v_1 v_2} + \tau_1 T^{walk}_{v_2 D} \qquad (20)$$

where $O$ and $D$ are user origin and destination, respectively. $v_1$ and $v_2$ are pickup and drop-off stops for user $s$ and route $r$, respectively. $\tau_1$, $\tau_2$ and $\tau_3$ are the value of walking time, value of waiting time and VOT, respectively. We set $\tau_1 = 1.5\tau_3$ and $\tau_2 = 2\tau_3$ (Wardman, Chintakayala, and de Jong., 2016). $\tau_3$ is 24.21 euro/hour as aforementioned.

The calibration result is shown in Figure 5. We vary $U_s^0$ from 0 to 100 and solve the P1 problem to match users and routes. We found $U_s^0 \geq 45$ euros fits observed user-route matches with 79.03% user-route matching rate on the training data based on the average of 5 runs. We retain $U_s^0 = 45$ as the calibrated constant user's trip utility value. For the remaining 20% test data, its corrected prediction rate of user-route matches is 65.45%. The mean and standard deviation of $U_{sr}$ is 73.39 and 3.57 for Belgium-side rides, and these numbers become 72.96 and 8.75, respectively, on the French-side. The standard deviation reflects the degrees of variation in $\hat{c}_s$ on these two areas.

We further test the normality assumption of $U_{sr}$. The skewness and kurtosis test for Normality shows the distribution of $U_{sr}$ for Belgium-side follows the normal distributions with p-value $(p > \chi^2) > 0.05$. For French-side, the Normality test is unable to be conducted due to its small sample size.

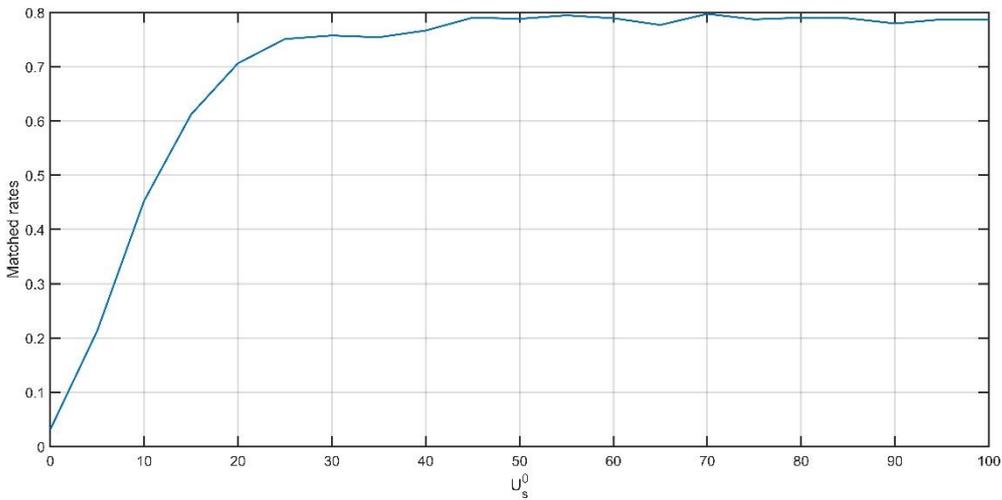

**Figure 5**. The calibration of constant part $U_s^0$ of trip utility.

## 3.4. Calibration of α

As we only have observed rides in the data and not on other modes the users may have taken, we calibrate the reliability parameter $\alpha$ based only on observed rides to fit model prediction and observations. The choice of $\alpha$ depends on the platform. When observing a platform as a third-party and given a distribution for the utilities, one can fit a value of $\alpha$ that maximizes the corrected prediction rate of the route flows matches. A two-stage computational procedure for calibrating the $\alpha$ for an $\alpha$-stable assignment game with heterogeneous users is described in Algorithm 1.

We use the training data set (first 80% rides (213 commuting periods)) for calibrating $\alpha$. For each commuting period $h \in H$, we have observed users and flows on Kussbus routes. We solve P1H and P2H for a given value of $\alpha$ to set up ticket price of users for each commuting period. Then we draw user's random ride utilities and insert the ticket price into P1 and solve P1 again to obtain the predicted route flows. We measure the corrected prediction rate over the training data set based on the difference between number of matches from the model and that from observed in the training data set, as shown in Eq. (21).

$$w_\alpha = 1 - \frac{\sum_{h \in H} \sum_{s \in S_h} |\hat{x}_{sr}^h - 1_{s\bar{r}}^h|}{\sum_{h \in H} |S_h|} \tag{21}$$

where $S_h$ is the set of observed rides (users) in the commuting period $h$. $1_{s\bar{r}}^h$ is an indicator being 1 if user $s$ uses route $\bar{r}$ in period $h$, and 0 otherwise. $\hat{x}_{sr}^h$ is the model prediction whether s uses the route $\bar{r}$ or not in period $h$.

From the average $\bar{w}_\alpha$ based on $K$ runs, we set a stable reliability parameter $\alpha$ with highest $\bar{w}_\alpha$. Afterwards, we can apply the model to other scenarios to anticipate how the platform would respond given their inferred $\alpha$.

Algorithm 1. Two-stage computational process of the stable matching model.

| | |
|---|---|
| 0: | Input: a set of candidate routes $r \in R$ and a set of observed rides, $s \in S_h$ over $|H|$ commuting periods, $h = 1,2,...,|H|$. Calibrate user's ride utility distribution (see Section 3.3) and compute user's generalized travel cost $t_{sr}$. |
| 1: | Set iteration $i = 0$, $\alpha_i = 0$, and step size $\Delta$. |
| 2: | While $\alpha_i \leq 0.5$ |
| 3: |   For $k = 1:K$ |
| 4: |     For $h = 1:|H|$ |
| 5: |       Solve P1H by leaving ticket price out and obtain the solution $\boldsymbol{x}_{sr}^h$. |
| 6: |       Given $\boldsymbol{x}_{sr}^h$ and $\alpha_i$, solve P2H based on user-optimal policy, i.e. $Z = \sum_{s \in S} u_s$, to maximize ridership and obtain the solution $u_s$ and $v_r$. Set up ticket price $p_{rs}$ by (19). |
| 7: |       Introduce $p_{rs}$ in Eq. (1), i.e. $a_{sr} = \max(0, U_{sr} - t_{sr} - p_{sr})$, and **solve P1** again to get the predicted route flows $\hat{x}_{sr}^h$. |
| 8: |     end |
| 9: |     Compute the corrected prediction rate over the training data set $w_\alpha^k$ by (21). |
| 10: |   end |
| 11: |   Compute $\bar{w}_\alpha = (w_\alpha^1 + \cdots + w_\alpha^K)/K$. |
| 12: | Set $\alpha_{i+1} = \alpha_i + \Delta$. Set $i := i + 1$ and go to step 3. |
| 13: | Retain best $\alpha^*$ with highest average corrected prediction rate $\bar{w}_\alpha$. |

Remark: 0.5 reflects the fact that we are interested in cases where the solution (user-route matches) is stable with probability higher than 0.5.

Note that we set up ticket prices based on the equal-share policy given the outcome obtained by P2. For example, consider a route $r$ with an operating cost of 40 euros and shared by 5 users. The portion of the payoff allocated to route $r$ from the solution of P2 is 20 euros. Under the equal-share policy, ticket prices for route $r$ are calculated as 40/5+20/5=12 euros. As aforementioned, $\alpha$ represent a reliability measure for which matches are perceived to be stable with the probability of $1 - \alpha$. We are interested in calibrating $\alpha$ within a set of discrete values, i.e. $\alpha \in (0.05, 0.1, 0.2, 0.3, 0.4, 0.5)$ to maximize the model prediction with the observed ridership. In practice, a platform or operator can further calibrate $\alpha$ within the range $0<\alpha \leq 0.5$ with higher precision. The calibration result is shown in Figure 6. We found $\alpha = 0.2$ has the best-fit of user-route matches with the average corrected prediction rate of 63.45%.

Table 6 reports the result of the stable matching model for the training and test dataset. For the training dataset, 76.38% of ride requests match Kussbus's operating routes, given $\alpha = 0.2$. For the test dataset, its average user-route matches are 70.11% with a 54.43% corrected prediction rate of route flows.

**Table 6. User-route matching result of Kussbus rides for 235 periods.**

| Data | Number of ride requests (users) | Number of user-route matches | User-route matching rate | Number of rides matched with observations | Matched rate (observation v.s. prediction) | Computational time (second) |
|---|---|---|---|---|---|---|
| Training dataset (80% obs.) | 2395 | 1829 | 0.7638 | 1520 | 0.6345 | 84.7 |
| Test dataset (20% obs.) | 615 | 431 | 0.7011 | 335 | 0.5443 | 29.0 |

Remark: $\alpha = 0.2$. The reported result is the average based on 5 runs.

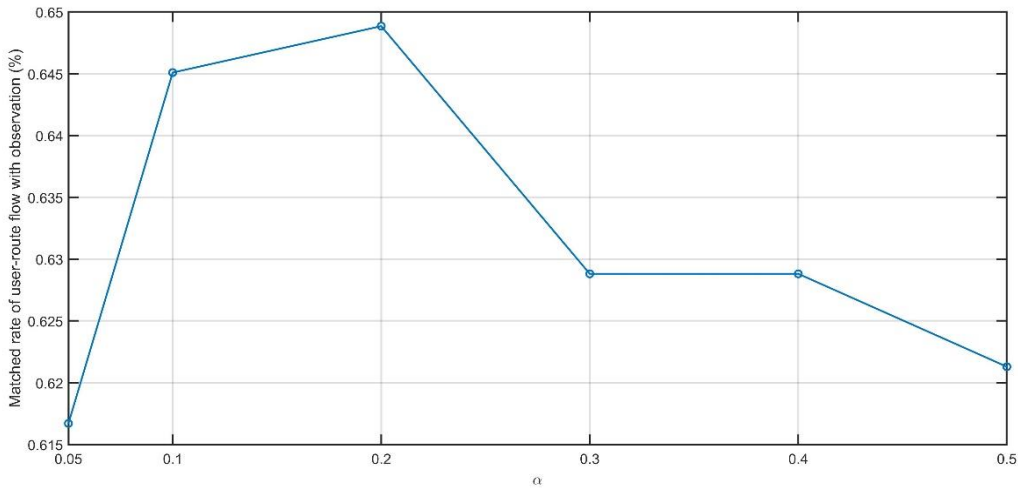

**Figure 6**. Corrected prediction rates of observed user-route matches for the training dataset over different $\alpha$.

## 4. Result

### *4.1. Detailed breakdown of two commuting periods using the stable matching model*

We now illustrate the detailed result of the stable matching model by considering two commuting periods on 06/27/2018 (Luxembourg-> Arlon in the evening) and 06/28/2018 (Arlon->Luxembourg in the morning) (see Table 7). There are 9 and 14 rides observed during these two periods. For the first period, 5 routes are matched with 9 users of which four routes are observed in the data. Only one route is different (Figure 7). The average ridership is 1.8 users/vehicle under the buyer-optimal policy. The ticket price ranges from 18.8 euros to 46.1 euros to ensure route operating cost could be covered from its revenue. As a comparison, when setting ticket prices under the seller-optimal policy, it would result in higher ticket prices for shared-ride users compared to that based on the buyer-optimal policy. For the second period, 4 routes are matched with 14 users which are observed to be identical (Figure 8). The average ridership is 3.5 users/vehicle with ticket price ranging from 9.2 euros for 6-users share and 28.8 euros for 2-users share. Figures 6 and 7 illustrates the detail of the spatial distribution of users' origins, destination and the operated routes based on observation and the model prediction.

**Table 7. Example of detailed result of the user-route matching model.**

| Period | Number of users | Route attributes | Assigned routes | | | | |
|---|---|---|---|---|---|---|---|
| 06/27/2018 Afternoon (Luxembourg->Arlon) | 9 | ID | 235 | 238 | 178 | 236 | 237 |
| | | Operating cost | 56.4 | 46.1 | 41.4 | 51.3 | 52.7 |
| | | Number of users | 3 | 1 | 1 | 2 | 2 |
| | | Ticket price ($p_{sr}$): | | | | | |
| | | →Buyer-opt. | 18.8 | 46.1 | 41.4 | 25.7 | 26.4 |
| | | →Seller-opt. | 47.1 | 50.1 | 48.8 | 55.6 | 42 |
| 06/28/2018 Morning (Arlon->Luxembourg) | 14 | ID | 242 | 240 | 239 | 241 | |
| | | Operating cost ($C_r$) | 57.6 | 55.2 | 52.1 | 54.1 | |
| | | Number of users | 2 | 6 | 3 | 3 | |
| | | Ticket price ($p_{sr}$): | | | | | |
| | | →Buyer-opt. | 28.8 | 9.2 | 17.4 | 18.0 | |
| | | →Seller-opt. | 49.7 | 50.0 | 46.4 | 50.4 | |

Remark: Ticket price and profit are measured in euros.

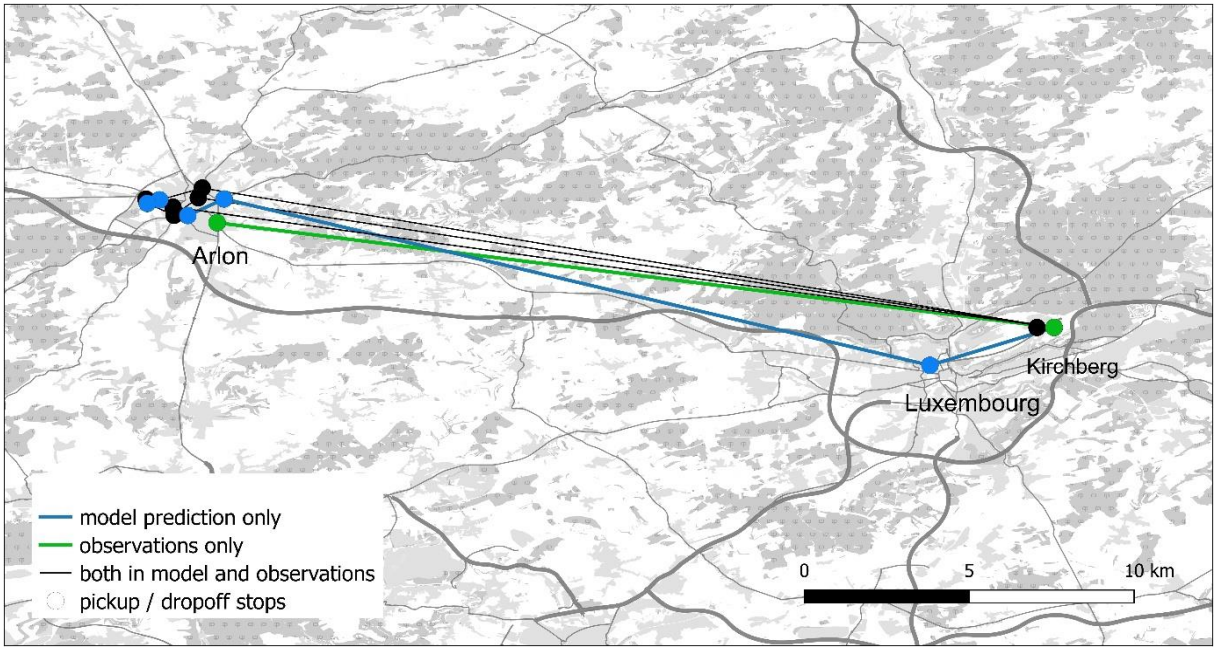

**Figure 7**. User-route match results of the stable matching model (Luxembourg to Arlon, 06/27/2018).

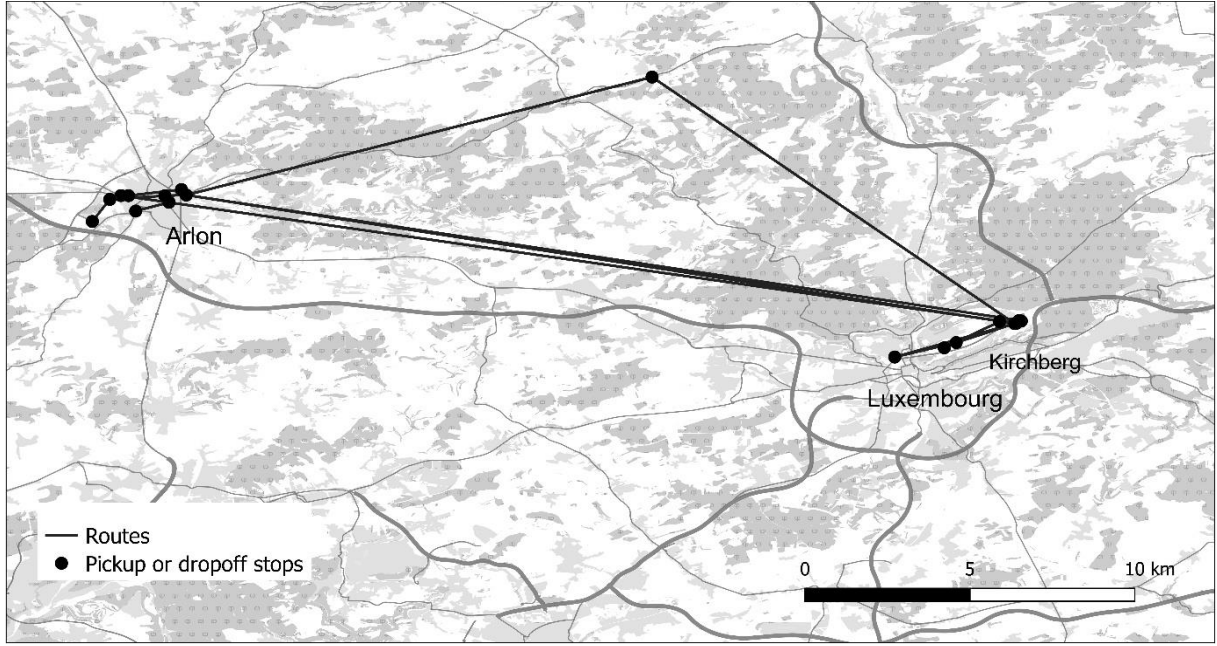

**Figure 8**. User-route match results of the stable matching model (Arlon to Luxembourg, 06/28/2018).

### 4.2. Comparison of the Kussbus pricing policy to the buyer-optimal and seller-optimal policies

We compare the result of the stable matching model based on buyer-optimal (i.e. $Z = \sum_{s \in S} u_s$ in (9)) and seller-optimal (i.e. $Z = \sum_{r \in R} v_r$ in (9)) cost allocation policies using calibrated $\alpha$ and the test dataset. The CDFs under different pricing policies are shown in Figure 9. For the buyer-optimal policy, the 50- percentile of the ticket price is 10.98 euros, and the 75- percentile is 13.18 euros. However, for the seller-optimal policy, a user's ticket price becomes 49.91 euros and 52.02 euros for the 50- and 75- percentiles, respectively. Compared to taxi fare[3] in Luxembourg (i.e. 2.5 euros for the initial charge and 2.6 euros per kilometer traveled), a single-ride Kussbus price is much cheaper compared to the current taxi fare. Note that Kussbus operated pricing policy gave 6 free rides to users and then charge around 5 euros per ride. Given no subsidy, the total revenue from its service operation is unable to compensate its total operating cost.

The total revenue, route cost and profit of the operator over the test dataset is shown in Table 8. The result is obtained from solving the stable matching model based on the four pricing schemes: Kussbus-operated ticket price, buyer-optimal ticket price, seller-optimal ticket price, and taxi fare. We find that Kussbus' operated policy would accumulate a financial loss up to -4135 euros for 465 matched users due to its lower ticket price compared to its route operating cost. By setting ticket prices based on the buyer-optimal policy, 426 users should match with the routes with a positive profit of 187 euros. By contrast, setting ticket prices based on the seller-optimal policy results in a relatively high ticket price (see Figure 9) due to the high operating cost (i.e. 61.0 euros/route on average, see Table 4). Consequently, only 6 users are matched with routes with a positive profit of 128 euros. Note that the counter intuitive result of why the seller-optimal case ends up with lower net profit is due to integrating the higher seller-optimal ticket price in the disutility function as explained in Section 2. Again, applying a taxi tariff results in no rides, given the high taxi fare for the long commuting distances of users in the studied area (i.e. average taxi fare is 114.3 euros, given an average travel distance of users is 43 km, see Table 4). We conclude that the buyer-optimal cost allocation policy is preferred to maximizing ridership and keeping the service at a minimum profitable level over the long term.

---

[3] https://www.bettertaxi.com/taxi-fare-calculator/luxemburg/

**Table 8. Revenue, operating cost and profit of different pricing policies for the test dataset.**

| Policy | Ridership | | Revenue | Operating cost | Net profit |
| --- | --- | --- | --- | --- | --- |
| Kussbus's tariff | 465 | (75.54%) | 1266 | 5401 | -4135 |
| Buyer-optimal ticket price | 426 | (69.24%) | 4831 | 4644 | 187 |
| Seller-optimal ticket price | 6 | (0.98%) | 231 | 103 | 128 |
| Taxi | 0 | (0.00%) | 0 | 0 | 0 |

Remark: Measured in euros. The reported result is the average of 5 runs.

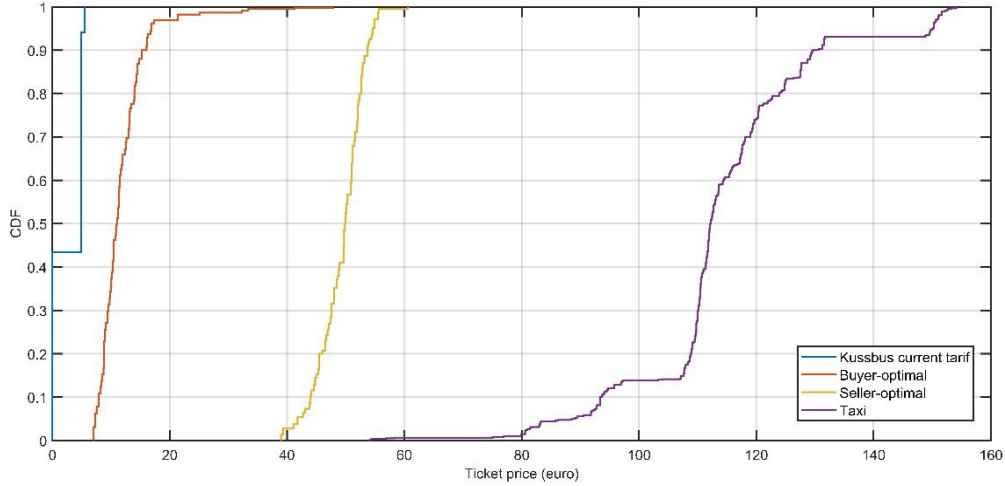

**Figure 9**. The cumulative probability distribution of user's ticket prices under different pricing policy for the test dataset.

### 4.3. Sensitivity analysis of policy

We further evaluate different system parameters to provide useful information for the operator to improve its operating policy design in the future. The considered decision parameters and the test scenarios are as follows.

- Scenarios 1: Route operating cost reduction: -10%,-30%,-30%,-40%,-50%. Examples of route operating cost changes include improvements in routing, repositioning, and matching algorithms that save idle time of vehicles, setting of common meeting points to streamline routes serving passengers, or reduction in congestion that leads to improvements in travel speed.
- Scenarios 2: In-vehicle travel time reduction: -10%,-30%,-30%,-40%,-50%. Examples include reduction in congestion leading to improvements in travel times for passengers.
- Scenarios 3: Access distance to bus stops reduction: -10%,-30%,-30%,-40%,-50%. Examples include algorithms that bring vehicles closer to travellers and reduce their access time.

Two ticket-pricing policies based on the buyer-optimal ($Z = \sum_{s \in S} u_s$) and seller-optimal ($Z = \sum_{r \in R} v_r$) setting are evaluated. The aim is to demonstrate the sensitivity of the model to the impact of different decision parameters on the ridership and profit of the operator.

We run the stable matching model based on the test dataset for different scenarios. The ticket price changes under different scenarios as shown in Table 9. For scenario 1, we find reducing route cost is most beneficial for users with lower ticket prices under the buyer-optimal policy. When reducing from -10% to -50% of the route cost, the ticket price would reduce from -6.9% to -37.9%. However, under the seller-optimal policy, the ticket price would keep stable with less than 1% variation.

**Table 9. Ticket price variation based on different scenarios.**

| Scenario (Reduction) | Route cost reduction scenario | | | | In-vehicle travel time reduction scenario | | | | Access distance to bus stops reduction scenario | | | |
|---|---|---|---|---|---|---|---|---|---|---|---|---|
| | BO | | SO | | BO | | SO | | BO | | SO | |
| | € | ±% | € | ±% | € | ±% | € | ±% | € | ±% | € | ±% |
| 0% | 11.6 | | 49.1 | | 11.7 | | 49.3 | | 11.6 | | 49.6 | |
| -10% | 10.8 | -6.9 | 49.4 | 0.6 | 11.7 | 0.0 | 51.1 | 3.7 | 11.7 | 0.9 | 49.4 | -0.4 |
| -20% | 10 | -13.8 | 49.4 | 0.6 | 11.8 | 0.9 | 52.7 | 6.9 | 11.8 | 1.7 | 49.9 | 0.6 |
| -30% | 9 | -22.4 | 49.3 | 0.4 | 11.8 | 0.9 | 53.9 | 9.3 | 11.4 | -1.7 | 49.9 | 0.6 |
| -40% | 8.2 | -29.3 | 49.1 | 0.0 | 12.2 | 4.3 | 55.1 | 11.8 | 11.6 | 0.0 | 50.0 | 0.8 |
| -50% | 7.2 | -37.9 | 48.9 | -0.4 | 11.9 | 1.7 | 55.7 | 13.0 | 11.3 | -2.6 | 50.3 | 1.4 |

Remark: BO: Buyer-optimal; SO: Seller –optimal. The result is based on the average of 5 runs.

For scenario 2, we find there is little change (less than 5%) observed for ticket prices based on the buyer-optimal policy. This is because the operator's route cost estimation depends on vehicle travel distance only. More elaborate route cost estimation that considers vehicle travel time can be integrated in the future. However, under the seller-optimal policy, reducing in-vehicle travel time between -10% to -50% would increase ticket prices between 3.7% to 13%. This is because the savings in travel time are absorbed by the operator in a seller-optimal policy.

For scenario 3, only a marginal variation (less than 3%) of the ticket price is observed for both pricing policies. As more than 95% of access distance to Kussbus bus stops is less than 1 km, it is expected that reducing the access distance further would have an insignificant impact on ticket prices. Figure 10 shows the cumulative probability distributions of ticket prices for different scenarios based on the buyer-optimal and seller-optimal policies.

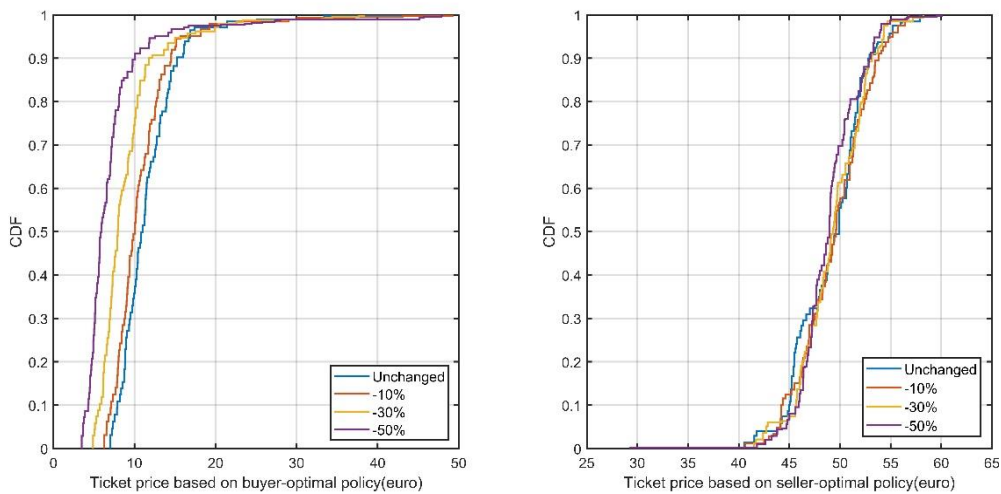

**Figure 10**. Influence of route cost reduction on ticket prices based on the buyer-optimal (on the left) and the seller-optimal policies (on the right).

*4.4. Policy recommendations under buyer- and seller-optimal policies*

The impact of different scenarios on ridership and the operator's profit under the buyer-optimal ticket price is shown in Table 10 and Figure 11. We found that reducing route cost is more effective to increase the ridership (up to +10% when 50% reduction of route cost) compared to reducing in-vehicle travel time and access distance to bus stops. However, it is not beneficial for the operator. For scenario 2, reducing in-vehicle travel time would slightly increase user-route matches (less than 2%) given fixed ride requests. However, it significantly increases the profit of the operator (i.e. +151.5% profit for 50% in-vehicle travel time reduction). Our study provides a benchmark under fixed demand. For the future extension, it would be interesting to consider flexible travel demand under a multimodal transport market setting. For scenario 3, a marginal impact on ridership and operator's profit is observed due to the short access distance to bus stops. As a conclusion of this comparison, we recommend government subsidy in support of scenario 1 while requiring Kussbus to operate under a buyer-optimal policy, as funding improvements in routing algorithms can significantly improve the consumer surplus of travellers.

**Table 10. Influence of different scenarios on the ridership and profit of the operator based on the buyer-optimal ticket price.**

| Scenario (Reduction) | Ridership | | | | | | Profit | | | | | |
|---|---|---|---|---|---|---|---|---|---|---|---|---|
| | S1 | | S2 | | S3 | | S1 | | S2 | | S3 | |
| | # | ±% | # | ±% | # | ±% | Euro | ±% | Euro | ±% | Euro | ±% |
| 0% | 431 | | 430 | | 430 | | 202 | | 196 | | 214 | |
| -10% | 428 | -0.7 | 419 | -2.6 | 431 | 0.2 | 199 | -1.5 | 142 | -27.6 | 187 | -12.6 |
| -20% | 436 | 1.2 | 431 | 0.2 | 447 | 4.0 | 129 | -36.1 | 276 | 40.8 | 236 | 10.3 |
| -30% | 460 | 6.7 | 458 | 6.5 | 436 | 1.4 | 113 | -44.1 | 304 | 55.1 | 92 | -57.0 |
| -40% | 467 | 8.4 | 433 | 0.7 | 434 | 0.9 | 148 | -26.7 | 373 | 90.3 | 119 | -44.4 |
| -50% | 474 | 10.0 | 438 | 1.9 | 436 | 1.4 | 12 | -94.1 | 493 | 151.5 | 161 | -24.8 |

Remark: S1: Route operating cost reduction scenario, S2: In-vehicle travel time reduction scenario, S3: Access distance to bus stops reduction scenario. The result is based on the average of 5 runs.

For the seller-optimal policy, we find reducing route cost and in-vehicle travel time could significantly increase both the ridership and profit of the operator compared to its benchmark as shown in Table 11 and Figure 12. Low ridership for the benchmark results from higher ticket prices. Under scenario 1 and 2, the number of rides would increase from initial 2 rides (over 615 requests) to 61 (scenario 1) and 72 (scenario 2). For scenario 3, its effect on the ridership and profit of the operator is less significant compared to the other two scenarios. In conclusion, if Kussbus were to operate on its own without government intervention, it can seek a seller-optimal policy and invest in algorithms that improve operating cost and/or in-vehicle travel time for passengers.

Our sensitivity analysis shows how the proposed stable matching model can be applied to evaluate different service designs. The operator can apply this methodology to set up ticket prices by considering the price ranges from buyer-optimal and seller-optimal policies.

**Table 11.** Influence of different scenarios on the ridership and profit of the operator based on the seller-optimal ticket price.

| Scenario (Reduction) | Ridership | | | | | | Profit | | | | | |
|---|---|---|---|---|---|---|---|---|---|---|---|---|
| | S1 | | S2 | | S3 | | S1 | | S2 | | S3 | |
| | # | ±% | # | ±% | # | ±% | Euro | ±% | Euro | ±% | Euro | ±% |
| 0% | 2 | | 2 | | 3 | | 81 | | 71 | | 85 | |
| -10% | 2 | 0 | 4 | 100 | 6 | 100 | 84 | 3.7 | 145 | 104.2 | 160 | 88.2 |
| -20% | 12 | 500 | 15 | 650 | 6 | 100 | 441 | 444.4 | 590 | 731.0 | 190 | 123.5 |
| -30% | 24 | 1100 | 33 | 1550 | 7 | 133.3 | 966 | 1092.6 | 1284 | 1708.5 | 229 | 169.4 |
| -40% | 38 | 1800 | 39 | 1850 | 5 | 66.7 | 1549 | 1812.3 | 1610 | 2167.6 | 151 | 77.6 |
| -50% | 61 | 2950 | 72 | 3500 | 10 | 233.3 | 2594 | 3102.5 | 3063 | 4214.1 | 394 | 363.5 |

Remark: S1: Route operating cost reduction scenario, S2: In-vehicle travel time reduction scenario, S3: Access distance to bus stops reduction scenario. The result is based on the average of 5 runs.

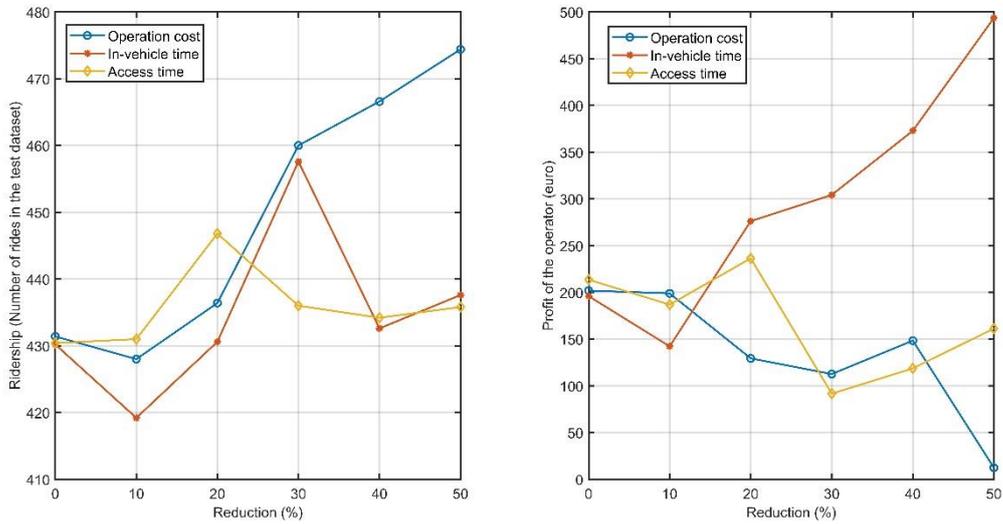

**Figure 11**. Influence of different scenarios on (a) ridership and (b) profit of the operator based on the buyer-optimal ticket price.

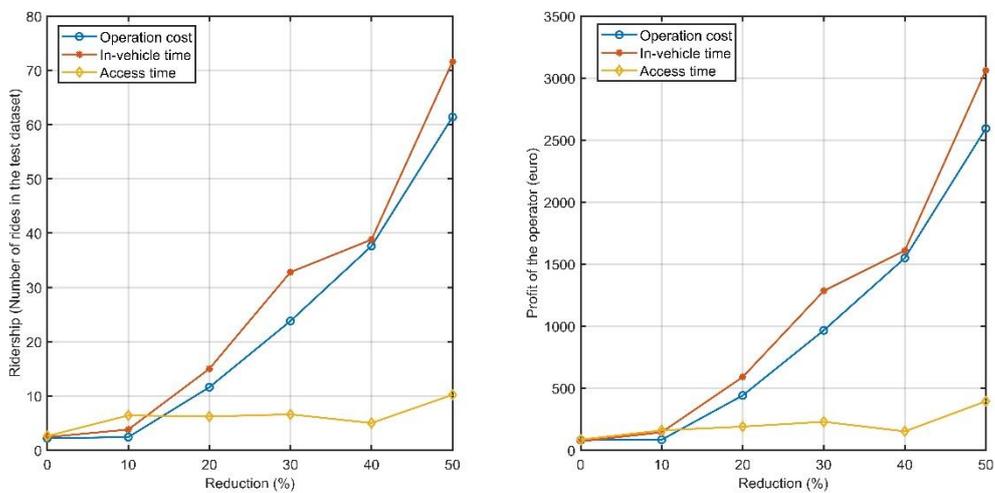

**Figure 12**. Influence of different scenarios on the ridership and profit of the operator based on the seller-optimal ticket price.

# 5. Conclusions

We tackle the problem of evaluating and designing a microtransit service. Microtransit operators can allocate resources to improve upon many aspects of operation: vehicle capacities; fleet size; algorithms to improve routing, pricing, repositioning, matching; and more. We conduct the first empirical application of a model from Rasulkhani and Chow (2019) that evaluates such systems using stable matching between travellers and operator-routes. The study is conducted using a real data set shared by industry collaborator Kussbus covering 3010 trips made between April to October 2018 in Luxembourg and its French-side and Belgium-side border areas.

In order to make it work empirically, we made several modifications to the model, primarily a conversion of the utility $U_{sr}$ into a function of different components including stochastic variables and making the stable matching model into a stochastic model. Doing so allows us to better fit the model to the data using the concept of $\alpha$-reliability.

We calibrated the model to the data. A separate data set from a mobility survey conducted in October to November 2012 covering a similar study area was used to estimate the value of time of travellers as 24.67 euros/hour, which we found consistent with existing VOT studies in Luxembourg. A base utility constant was then estimated for travellers in the Kussbus data and found to be 45 Euros to obtain 79.03% matching rate with the training data. Validation using the 20% test data showed a user-route match rate of 65.45%. The value of $\alpha$ was calibrated to a value of 0.20 as the best fit to the observations with a corrected prediction rate of 63.45% resulting in 76.38% ride matches. Validation with the 20% test set resulted in 54.43% corrected prediction rate with 70.11% matches.

Our stable matching model, as illustrated with two commuting periods, shows the existence of a stable outcome space between buyer-optimal and seller-optimal policies. We show that Kussbus current pricing policy falls below the buyer-optimal policy, which is not sustainable. By increasing the ticket price to the buyer-optimal policy it would reduce ridership from the current 465 trips to 426 trips and changing the net profit from -4135 euros to 187 euros for 615 ride requests. Increasing the pricing allocation further to the seller-optimal policy significantly reduces the ridership and reduces net profit, while following taxi pricing policy would lead to zero trips.

A sensitivity analysis is then conducted to compare the effects that equal, unilateral reductions in route operating cost, in-vehicle travel time, and access distance to bus stops, can have on the microtransit service. We find that government can intervene by offering to subsidize Kussbus to improve their routing algorithms and reduce operating cost while requiring operation under a buyer-optimal policy. Such an intervention can increase ridership by 10% with an operating cost reduction of 50%. Alternatively, an independent Kussbus can operate in a seller-optimal policy and invest in algorithms to improve in-vehicle travel time which can improve profit by 731% (bearing in mind the low ridership if operating a seller-optimal policy in the current baseline setting) with a 20% reduction in in-vehicle travel time. These analyses can be further conducted with other operational variables like fleet size, fleet mix in vehicle size, service coverage area, and more.

New insights have been made as a first empirical study of microtransit operation using the stable matching modelling framework. However, more research can be done to improve this work further. A study that includes travellers as part of a whole market system would capture their utility preferences better, allowing us to specify choice models and using the utility functions for the stable matching model. Alternatively, methodological extensions can be made to allow us to evaluate platforms (see Chapter 3.5 in Chow, 2018) controlling submarkets in the presence of external operators/platforms. Evaluation of the Kussbus service as a potential component of a multimodal MaaS trip (see Pantelidis, Chow and Rasulkhani, 2019) would be a much more powerful study that can relate its operational policies to impacts to the MaaS market. In this study, a static multi-period model is used to fit to the data; a more realistic model would be a dynamic model that considers dynamic cost allocations (e.g. Furuhata et al., 2014).

**Acknowledgements**

The work was supported by the Luxembourg National Research Fund (INTER/MOBILITY/17/11588252), National Science Foundation grant CMMI-1634973, and C2SMART University Transportation Center. We thank the Utopian Future Technologies S.A. (UFT) for providing Kussbus riding data for this research. Particular thanks to Ms. Kimberly Clement for her technical support for the empirical data analysis and to Saeid Rasulkhani for providing the computer program from the previous study and helpful comments.